\documentclass[conference]{IEEEtran}
\IEEEoverridecommandlockouts

\usepackage{microtype}
\usepackage{cite}
\usepackage{amsmath,amssymb,amsfonts}
\usepackage{algorithmic}
\usepackage{graphicx}
\usepackage{textcomp}
\usepackage{xcolor}
\usepackage{hyperref}
\usepackage{mwe,tikz}
\usepackage[percent]{overpic}
\usepackage{amsmath,amsfonts}
\usepackage{algorithmic}
\usepackage{graphicx}
\usepackage{textcomp}
\usepackage{xcolor}
\usepackage{color} 
\definecolor{blueviolet}{RGB}{138,43,226}
\usepackage{booktabs}

\makeatletter
\let\MYcaption\@makecaption
\makeatother

\usepackage[font=footnotesize]{subcaption}

\makeatletter
\let\@makecaption\MYcaption
\makeatother\usepackage{listings}

\usepackage{times}
\usepackage{pifont}
\usepackage{flushend}
\newcommand{\cmark}{\ding{51}}%
\newcommand{\xmark}{\ding{55}}%
\newcommand{\ourwork}[1]{\leavevmode\color{blueviolet}#1}

\usepackage[frozencache=true]{minted}
\usepackage{array,tabularx}
\newcolumntype{b}{X}
\newcolumntype{r}{>{\hsize=.3\hsize}X}
\newcolumntype{k}{>{\hsize=.20\hsize}X}
\newcolumntype{m}{>{\hsize=.50\hsize}X}
\newcolumntype{n}{>{\hsize=.60\hsize}X}
\newcolumntype{s}{>{\hsize=.15\hsize}X}

\newcommand{\final}{0} 
\newcommand{\muhammad}   [1]{{{\color{red}(muhammad) #1}}}
\ifthenelse{\equal{\final}{1}}{\renewcommand{\muhammad}[1]{}}{}
\newcommand{\ryan}   [1]{{{\color{purple}(ryan) #1}}}
\ifthenelse{\equal{\final}{1}}{\renewcommand{\ryan}[1]{}}{}
\newcommand{\karthik}   [1]{{{\color{green}(karthik) #1}}}
\ifthenelse{\equal{\final}{1}}{\renewcommand{\karthik}[1]{}}{}
\newcommand{\jalal}   [1]{{{\color{purple}(jalal) #1}}}
\ifthenelse{\equal{\final}{1}}{\renewcommand{\jalal}[1]{}}{}

\usepackage[most]{tcolorbox}

\tcbset{textmarker/.style={%
        enhanced,
        parbox=false,boxrule=0mm,boxsep=0mm,arc=0mm,
        outer arc=0mm,left=4mm,right=3mm,top=7pt,bottom=7pt,
        toptitle=1mm,bottomtitle=1mm,oversize}}

\newtcolorbox{api_box}{textmarker,
    borderline west={6pt}{0pt}{pastelyellow},
    colback=pastelyellow!10!white}

\newtcolorbox[auto counter]{sidebar_box}[2][]{textmarker,
    floatplacement=t,
    float,
    borderline west={6pt}{0pt}{pastelblue},
    colback=pastelblue!10!white,
    text width=\columnwidth,
    halign=justify,
    title=\textcolor{black}{\textbf{Sidebar~\thetcbcounter}~#2},
    title code={
      \path[fill=pastelblue!10!white] (title.south west) rectangle (title.north east);
      \path[draw=pastelblue,solid,line width=0.75mm]
      ([xshift=0mm]title.south west) -- ([xshift=0mm]title.south east);
      },
    nameref={#2},
    #1
}

\def\BibTeX{{\rm B\kern-.05em{\sc i\kern-.025em b}\kern-.08em
    T\kern-.1667em\lower.7ex\hbox{E}\kern-.125emX}}
\begin{document}
\title{Seer: Predictive Runtime Kernel Selection\\ for Irregular Problems}

\author{\IEEEauthorblockN{Ryan Swann}
    \IEEEauthorblockA{ryan.swann@amd.com\\
        AMD Research, USA}
    \and
    \IEEEauthorblockN{Muhammad Osama}
    \IEEEauthorblockA{muhammad.osama@amd.com\\
        AMD Research, USA}
    \and
    \IEEEauthorblockN{Karthik Sangaiah}
    \IEEEauthorblockA{karthik.sangaiah@amd.com\\
        AMD Research, USA}
    \and
    \IEEEauthorblockN{Jalal Mahmud}
    \IEEEauthorblockA{jalal.mahmud@amd.com\\
        AMD Research, USA}
}

\maketitle

\begin{abstract}
    Modern GPUs are designed for regular problems and suffer from load imbalance when processing irregular data. Prior to our work, a domain expert selects the best kernel to map fine-grained irregular parallelism to a GPU. We instead propose Seer, an abstraction  for producing a simple, reproduceable, and understandable decision tree selector model which performs runtime kernel selection for irregular workloads. To showcase our framework, we conduct a case study in Sparse Matrix Vector Multiplication (SpMV), in which Seer predicts the best strategy for a given dataset with an improvement of 2$\times$ over the best single iteration kernel across the entire SuiteSparse Matrix Collection dataset.
\end{abstract}


 \section{Introduction}
\label{sec:introduction}


Modern Graphics Processing Units (GPUs) have been proposed to leverage
the available parallelism in problems such as linear algebra, high-performance scientific computing, graph analytics, and more. Many workloads are
dense and regular, where all threads within the GPU receive equal amounts of parallel work to process. It is not always the case, however, that parallel workloads are regular in nature. Problems such as graph analytics and sparse-linear algebra have an abundance of available fine-grained but irregular parallelism~\cite{Wang:2017:GGG}.

The design of a GPU is fundamentally challenged when the volume of parallelism available is not regular or structured~\cite{Osama:2023:APM}. Within irregular problems each thread within a Single Instruction/Multiple Data (SIMD)~\cite{Flynn:1966:VHS} lane receives a different allotment of work. This imbalance leads to issues in which threads with less work may be waiting on threads with more work. This \emph{load imbalance} causes the efficiency of execution on irregular problems to be data dependent~\cite{Osama:2022:GLB}. This fundamental limitation of GPUs has inspired the adoption of many \emph{load balancing techniques}~\cite{Osama:2023:APM,Busato:2017:APP}
as well as compressed sparse formats~\cite{Filippone:2017:SMM}, which hope to extrapolate the best strategy and data-structure to map irregular data to the naturally regular graphics hardware.

\begin{figure}
  \centering
  \includegraphics[width=\linewidth]{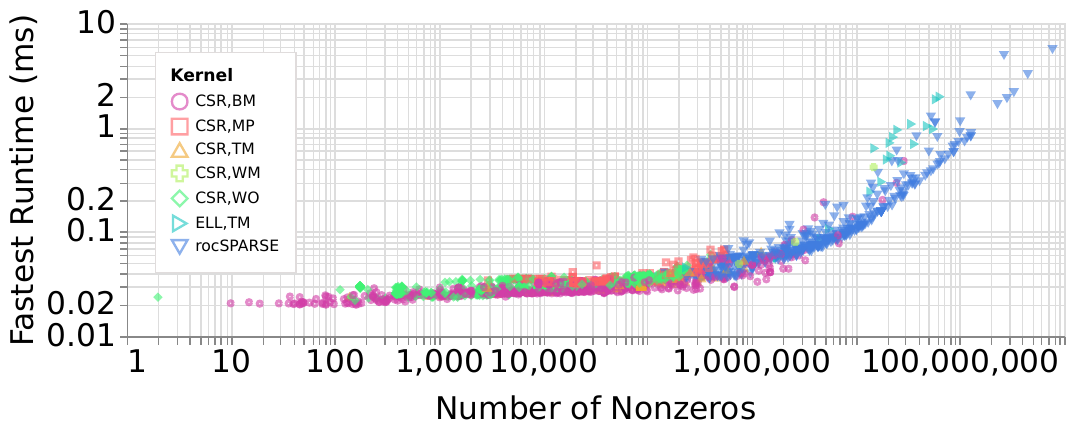}
  \caption{Kernel implementation with the best runtime (fastest) for a given dataset being selected. The wide range of most performant techniques (i.e., range of colors) for varying datasets motivates the need for a kernel selection predictor for irregular problems. \label{fig:lb_problem}}
\end{figure}

Due to the regular nature of the graphics processor architecture and
the broad range of different forms which data may come in, there is not an objective method
for technique or data structure selection which does not require explicit tuning by a
domain-expert. Thus, many libraries implement the load balancing technique which is
best for their workloads without consideration of other techniques which may yield
better results depending on the shape of the input data. Fig.~\ref{fig:lb_problem} illustrates this problem by visualizing performance of a Sparse-Matrix Vector Multiplication (SpMV) kernel with several different load-balancing schedules and compressed formats on the SuiteSparse Matrix Collection~\cite{Davis:2011:TUO}. The techniques and formats used are discussed in detail later in the paper. Each color and shape represent a different kernel being selected for a given dataset, and it shows that even for problems with similar amounts of work (a vertical slice of the plot), there can be a different kernel selection for optimal performance. Selecting the best load-balancing schedule and compressed sparse format at runtime without costly overhead is a challenging and non-trivial problem.

To address this problem, we propose \emph{Seer}, a machine learning-based predictor which considers the shape of the
irregular data at runtime and uses a light-weight machine learning model to provide an inference which selects the best method to map the irregular data to the hardware. The contributions of this paper are as follows:

\begin{enumerate}
  \item A predictive runtime framework and abstraction for kernel selection. Our framework considers the cost of preprocessing for iterative algorithms, the cost of feature collection, and various compressed sparse formats and load-balancing schedules, all built on an explainable decision tree model.
  \item A classifier selection predictor that weighs the cost of feature collection versus a no overhead approach and selects the best method of prediction at runtime.
  \item A case study of SpMV using our method yields 2$\times$ better performance over the best kernel and a geomean speed-up of 6.5$\times$ across the test set.
\end{enumerate}

\section{Background and Related Works}
\label{sec:background}

Past works have shown that the performance of a particular kernel may vary significantly based on the method that is used to map work to  threads or the corresponding data structure used for the implementation. At the fundamental level, the problem could be summarized as compute and memory optimizations targeted at various structured and unstructured patterns that emerge within the irregular dataset and algorithms. Filippone et al. \cite{Filippone:2017:SMM} and Busato et al. \cite{Busato:2017:APP} provide a detailed summary of compressed data structures and load-balancing algorithms used in the literature for one such irregular workload, SpMV.

While there are many proposed load balancing schedules~\cite{Osama:2022:GLB} and memory-system aware data structures~\cite{Filippone:2017:SMM} in literature to achieve high performance for irregular parallel problems on the GPU architecture, there is no formal definition for how to relate these techniques to any arbitrary dataset. The na\"ive approach is to run all permutations of load-balancing schedules and compressed sparse data structures that are implemented, and pick the best one (e.g., fastest runtime) for a given problem. However, this strategy is unrealistic for runtime execution. The contributions of other works which attempt to automate this process are summarized and compared to our framework, Seer, in Table~\ref{fig:comparison_table}. Works addressing ML for systems  note that ML-based prediction techniques are skilled at extracting features that may be implicit to humans without requiring explicit implementation of every case in code \cite{Wu_2022}.

\begin{table}[b]
    \centering
    \caption{Feature comparison of our work to prior works. \label{fig:comparison_table}}

    \begin{tabularx}{\columnwidth}{Xsrrr}
        \toprule
        Feature                    & \ourwork{\textbf{Seer}} & Nitro~\cite{Muralidharan:NITRO} & WISE~\cite{Yesil:WISE} & spECK~\cite{Parger:SPECK} \\
        \midrule
        Preprocessing Amortization & \ourwork{\cmark}        & \xmark                          & \xmark                 & \xmark                    \\
        Feature Collection Cost   & \ourwork{\cmark}        & \xmark                          & \xmark                 & \cmark                    \\
        Classifier Selection Model             & \ourwork{\cmark}        & \xmark                          & \xmark                 & \xmark                    \\
        General Abstraction        & \ourwork{\cmark}        & \cmark                          & \xmark                 & \xmark                    \\
        Sparse Case Study          & \ourwork{\cmark}        & \cmark                          & \cmark                 & \cmark                    \\
        Compressed Formats         & \ourwork{\cmark}        & \cmark                          & \cmark                 & \cmark                    \\
        Explainability             & \ourwork{\cmark}        & \xmark                          & \cmark                 & \xmark                    \\
        \bottomrule
    \end{tabularx}
\end{table}

Nitro~\cite{Muralidharan:NITRO} provided an autotuning framework for code-variants, which are effectively different kernels which implement the same function. Nitro provided a framework for the abstraction of these code-variants which a programmer could utilize, along with provided features, to select run variants using a Radial-Basis Function at runtime. Nitro~\cite{Muralidharan:NITRO} considers several case studies including SpMV, BFS, and sorting. Nitro's framework is not directly targeted at irregular workloads, but is able to function as a predictor for irregular workloads with reasonable performance. Nitro mentions that the cost of feature collection may be amortized over many iterations, but does not explicitly consider the cost of this collection at runtime.

WISE~\cite{Yesil:WISE}  focuses specifically on SpMV with a large focus on selection of representative features. WISE uses a large number of parameters in which the relationship between the features and the execution is concretely defined. WISE therefore provides its model a very complete representation of the input through its robust set of input features. WISE utilizes a decision tree model for their machine learning framework, which is provided with the extensive set of features for training, and is able to achieve 2.4x speedup over the Intel Math Kernel Library\textregistered \cite{Intel:MKL}  without considering the cost of feature collection.

spECK~\cite{Parger:SPECK} is a machine learning framework for dynamic parameter selection for Sparse Matrix-Matrix multiplication and considers the cost of gathering features in the design of their machine learning framework with different kernel variants. spECK is tailor-made for utilization in Sparse Matrix-Matrix multiplication and does not provide a generalization of the method. spECK features a modular kernel which can be actively configured to gather more or less data based on the input. Usage of spECK results in the fastest execution time on 79\% of plots in their dataset and second best on 15\%.

The past works largely address the concept of utilizing a machine learning model for irregular workloads, but several do not generalize their approach, which is important due to the large breadth of irregular workloads. Additionally, several other works do not consider the concept of preprocessing amortization \cite{Muralidharan:NITRO} \cite{Yesil:WISE}, which is a situation where the cost of an initial preprocessing stage on a sparse matrix may be amortized over several iterations. While single iteration performance is important for some irregular workloads, multi-iteration runs are also a very common use case in irregular workloads. The ability to predict preprocessing amortization can often significantly impact the performance for multi-iteration runs and therefore is an important consideration.

While past works have covered the importance of being able to select kernels, there are seldom mentions of the cost of collecting data which they then use for inference. The lack of attention to feature collection cost causes the formerly proposed solutions to be unrealistic for utilization in kernel selection at runtime. In many of the prior works \cite{Muralidharan:NITRO}\cite{Parger:SPECK}, the authors do not take into account the cost of feature collection, or account for it in passing, which for many statistics collection routines may overcome the cost of running the workload itself. This insight is especially true in instances where a single iteration of a workload is run, or the runtime of the workload itself is very low. In these instances, it is difficult to justify the cost of running additional feature collection kernels at all. In the next section, we show that Seer is not only amenable to this cost, but it is considered in the design of the abstraction.

\section{Abstraction and Framework}
\label{sec:framework}

Seer, our framework for predictive kernel selection is built on a two-level abstraction, one which focuses on the training of the model and another that deploys an efficient classifier selection model for runtime inference.
\begin{figure*}
    \begin{minipage}[t]{\textwidth}
        \includegraphics[width=\linewidth]{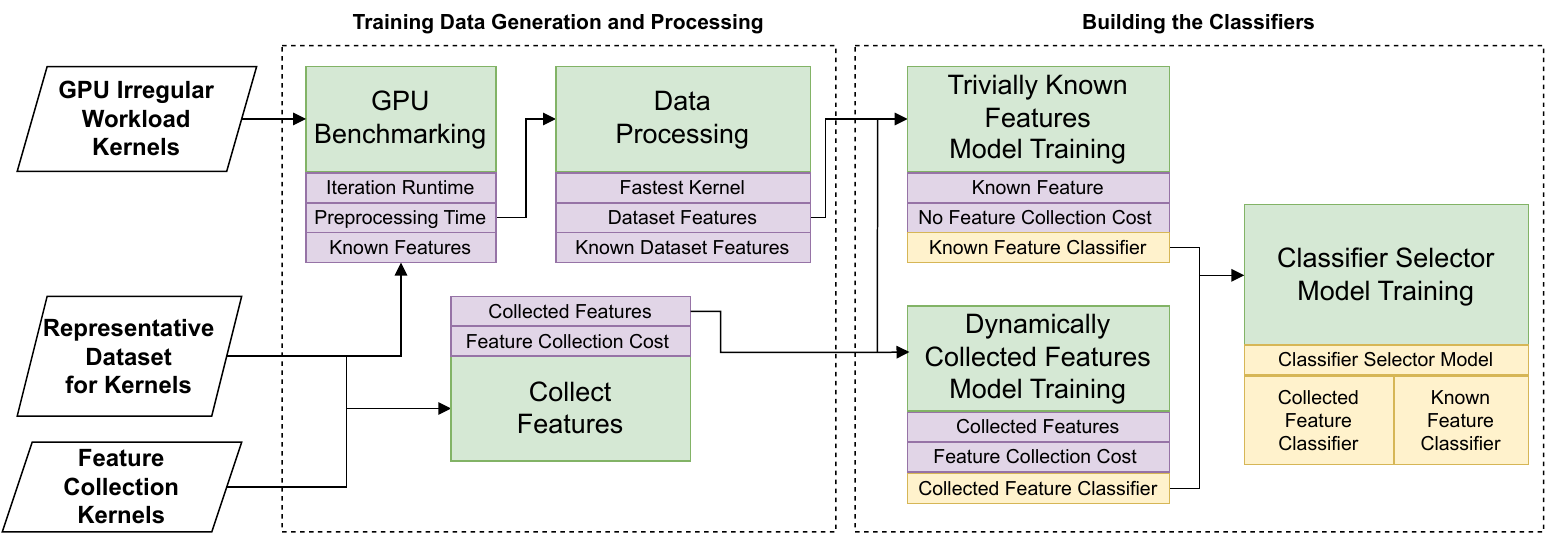}
        \caption{Training Abstraction showing the process for training three different models. A \emph{known features} classifier is trained on the trivially known features of the data given with the dataset. A \emph{gathered feature} classifier is trained on statistical data that is gathered with an additional cost. Both the \emph{known} and \emph{gathered} data classifiers are incorporated in an \emph{classifier selection} in which a third classifier is trained to select between the \emph{known} and \emph{gathered} models. \label{fig:training_abstraction}}
    \end{minipage}
\end{figure*}

\begin{figure}
    \centering
    \includegraphics[width=0.75\linewidth]{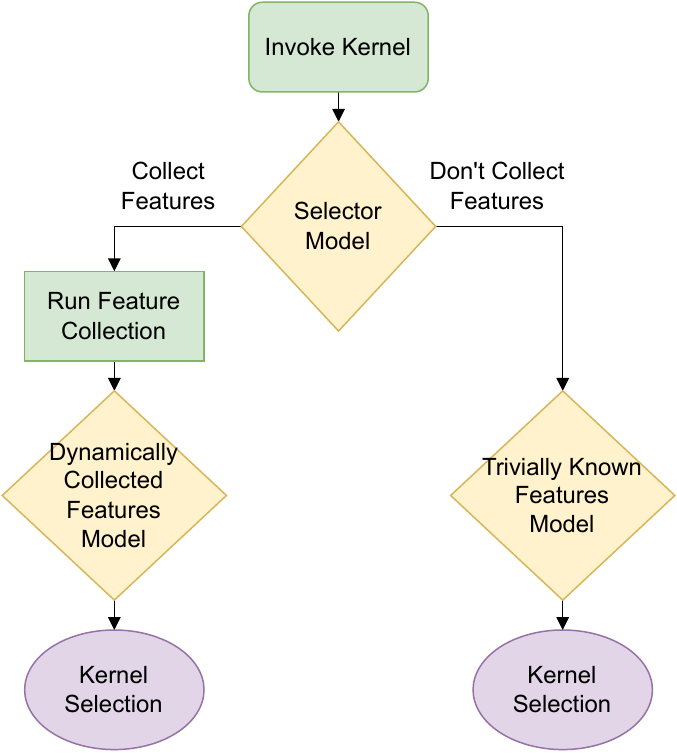}
    \caption{Inference process at runtime uses a \emph{classifier selection model} to select between the cost of dynamically computing data features and only utilizing trivially known features for prediction. \label{fig:inference_flowchart}}
\end{figure}


\pagebreak
\subsection{Training}

The training abstraction, as shown in Fig.~\ref{fig:training_abstraction}, requires three inputs:
\begin{itemize}
    \item \textbf{GPU irregular workload kernels}, the kernels of interest which exhibit irregular patterns and vary in performance depending on the shape of the data. The model consumes the performance metrics (e.g., runtime) as an output to be optimized.
    \item A \textbf{representative dataset}, which is identified as a dataset that is adequately representative of the work which the user actually plans to run.
    \item \textbf{Feature collection kernels}, which collect dynamically computed features from the representative data and output it to a file which is consumed by our framework.
\end{itemize}

\vfill\null


Our abstraction distinctly identifies the difference between \emph{known} and \emph{gathered} features\footnote{In the HPC context, these data features are often called metadata of the particular dataset, such as matrix dimensions, average row length, or metrics defining the structure of the sparsity.} when training. Within irregular datasets, some dimensionality of the data is known (e.g., rows and columns of a matrix). These metrics are considered \emph{known} features, as they are known at no additional runtime cost. \emph{Gathered} metrics, however, are collected at runtime through additional pre-processing steps. The collection of these dynamically computed features has a non-zero cost which must be considered as an overhead of the gathered-feature prediction model. We train decision trees using the CART algorithm implemented by scikit with Gini Impurity as the splitting criterion. To avoid overfitting, we implement a max depth on the decision tree and do not tune hyperparameters on the test set.
The type of features which should be utilized are entirely workload dependent. Due to this workload dependence, features are an input for which the user is able to provide a representative dataset and the parallel or sequential kernels that collect a set of statistics to create a representation of the data for training. These statistics (e.g., average row-length of a sparse-matrix) are then used within training to create a more accurate predictor for given kernels. 

We first train a classifier based on the trivially known features and task it with classifying the fastest kernel given this data. Due to this classifier's ability to predict solely on the trivially known features that are known at runtime, it incurs no additional overhead. As such, it is conjectured that this model is best applied in cases where an accurate prediction can likely be made on the trivially known features alone, and the cost of misprediction is low.
Next, we train a classifier model based on both the trivially known and dynamically computed features and task it with classifying which kernel will be the fastest given the known and gathered features. Due to this classifier's precondition of feature collection, it is conjectured that this model will be best utilized when a misprediction is costly, and the probability of a misprediction is high. This model achieves the highest accuracy, but at the cost of its feature collection precondition.
Given both the \emph{Known} and \emph{Gathered} feature models, it becomes necessary to make the decision between which of the two models to use at runtime. For this purpose, we introduce a classifier selection model which selects between the known and gathered feature models.

\subsection{Runtime Inference}
\label{sec:runtime_inference}

For the purposes of this paper, inference is considered to be executed at runtime. Within Seer, we consult the classifier selection model, which arbitrates which of the two models will be utilized for the inference process (see Fig.~\ref{fig:inference_flowchart}). The classifier selection model takes the trivially known features of the dataset as input and provides a classification which determines if more data features are required to make an accurate inference. It also determines if the cost of collecting the dynamically computed features is justified by the possible improvement in runtime from a correct prediction. The output of this model is a classification between one of the sub-models.

The inference in a decision tree is effectively a number of nested if-else statements which take in an input vector and then decide based on the internal weighting what output should be selected based on the input vector. This output is typically an integer which represents an index in the output vector.

Next, we run either the model based on both dynamically computed features and trivially known features or the model based on only the trivially known features, depending on the prediction of the classifier selection model. If the classifier selection model selects the gathered features model,  the program must gather the dynamically computed features at runtime before running the gathered feature model. To accomplish this, the program must first run the feature collection kernels, which will  provide the dynamically comptued features required for the gathered feature model. Once the dynamically computed features are gathered, they are passed into the gathered feature model which subsequently provides a classification as to which kernel it believes will give the fastest runtime for a dataset.
Similarly, if the known feature model is selected then there is no precondition of data collection which must be met. Given the trivially known features, we can  immediately make an inference to classify which kernel will perform best on a dataset based on the trivially known features.

\subsection{Design Decisions}

In this section we will cover the rationale behind some of the design decisions that were made in building the model. Firstly, a Classifier model was chosen due to the nonlinear relationship between the graph's shape characteristics and the runtime. We have also tested with linear regression and gradient boosting based methods before deciding on a classifier model. These other models which provide a quantitative output required significantly more information to make an accurate inference and were unable to capture the relationship between the data and a kernel's runtime. After these tests and a review of related works discussed in section \ref{sec:background}, a classifier model was selected as it was noted that the problem we were trying to solve is not runtime prediction but rather prediction of which category of kernel to select, which restricts the output space to the a selection between the number of kernels.

Specifically, we selected a decision tree classifier due to its negligible runtime overhead and its explainability. The decision tree is lightweight as it is effectively a set of if-else statements with weights that decide how to traverse the tree. This seems intuitively similar to the complex handwritten selection processes that have previously been used. In terms of explainability, the decision tree's weights may be output, and the decision tree may be viewed as a static piece of code with weights that do not change. By viewing the tree this way, we hope we can gain an intuition as to what exactly makes a kernel more performant and thus be able to explain why a particular kernel was selected. This also is more amenable to production libraries, as a chosen tree shall be deterministic, and thus a particular decision is traceable to a set of weights.

Another decision that may seem perplexing at first is the choice to use a classifier selection model of a known and gathered features model, especially as this is core to the abstraction. Initially, we took the simplest route and made the trivially known feature predictor which showed reasonable accuracy. To the accuracy, we found it necessary to provide the model with more information, but the collection of these dynamically computed features felt slow and intuitively did not seem make sense for a tool meant to be used at runtime. In attempting to parallelize the dynamic computation of these features, we recognized that the collection of the dynamically computed features in some instances resulted in a parallel feature collection runtime which is higher than the cost of running the sparse matrix, even for simple statistics. From this, we were able to recognize that it is unrealistic to collect dynamically computed features in many instances, and there should be a way to utilize the simple predictor and only resort to the gathering of dynamically computed features when it is necessary to do so. We found that the classifier selection model was capable of providing a classification which could make the selection between the known and gathered data predictors.

At training time, we wanted to avoid overfitting the model to the training data so we avoided hyperparameter tuning using the training set and set a maximum depth of the tree. Setting a maximum decision tree depth avoids overfitting in a decision tree as otherwise branches will continue splitting until they have 0 impurity, resulting in a perfect fit of the data. As we do not perform explicit hyperparameter tuning, we do not use a validation set.

Additionally, it is conjectured that the classifier selection model could be further extended to gather different types of information depending on an intial inference. That is, the classifier selector could become a selector of a larger number of models where each class of its output collects a different subset of the statistics. However, this problem is for future research.

\subsection{Seer API}

To implement our generalized abstraction for kernel selection for fine-grained parallel irregular computations, we provide a modular framework built on simple APIs. The Seer API is defined in the following stages: GPU benchmarking, feature collection, data processing, and the models.

\begin{figure*}
    \includegraphics[width=\linewidth]{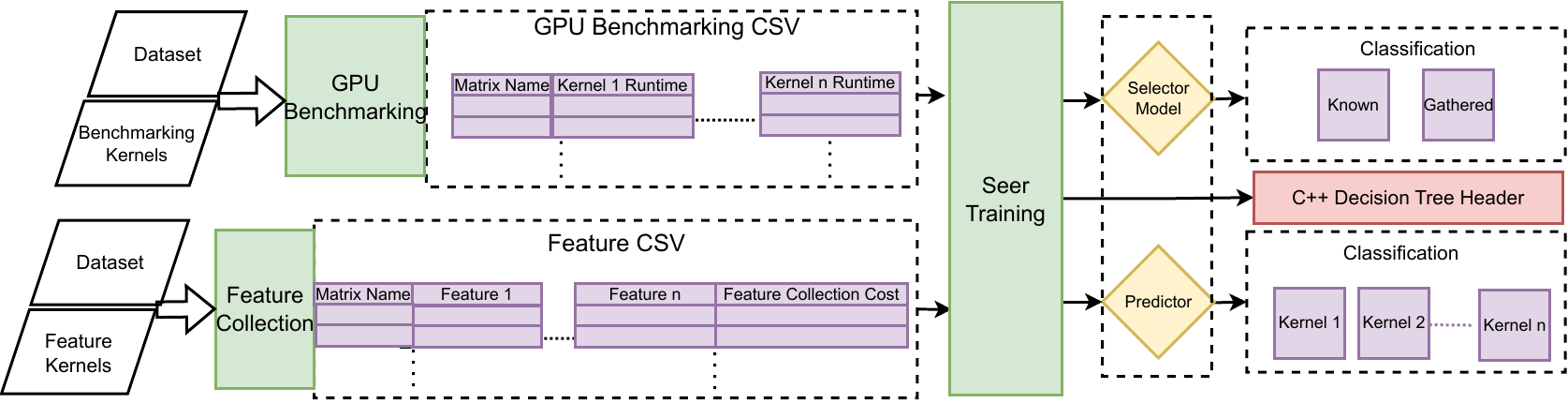}
    \caption{A high-level overview of the Seer API. First, the dataset and kernels are passed into the GPU benchmarking and feature collection kernels. These kernels output a CSV which is input to the Seer training script and used to generate the classifier selection model and its predictors. The predictors are output as both an object in the training script for further analysis as well as a C++ Header File which may be imported in C++ for usage.}
\end{figure*}

In the GPU benchmarking stage, the user provides a set of kernels as input and these kernels collect  features from the representative dataset provided by the user. The dataset is expected to be representative of the data the user would like to predict. The GPU benchmarking stage then outputs a CSV which contains three columns. These three columns are the name of the input, the runtime of the kernel, and the preprocessing time of the kernel. These files are produced separately for each kernel, and later aggregated into a single file. The output after this aggregation is two files with columns equal to the number of kernels plus 1. These final aggregate files have one column which represents the name of the data, or some unique identifier for a member in the dataset. Every other column contains a header with the name of the kernel followed by the runtime for each individual member of the dataset.

In the feature collection phase, the user provides as input a set of kernels which collects dyanamically computed from the representative dataset as well as the representative dataset. For each element in the representative dataset, the dynamically computed are collected in a CSV file which contains data features + 2 columns. The first column is a unique name for the member of the dataset, and the last column contains the collection time for the feature in the row. The columns in between contain the individual feature characteristics for the representative dataset with one column each.

After both the dynamically collected features and runtime information are collected, the data is passed into the Seer training script to be processed by calling:
\begin{minted}[fontsize=\small,
        obeytabs=true,
        tabsize=4,
        linenos=false,
        % frame=lines,
        escapeinside=||,
        numbersep=-1pt]{c++}
  seer(runtime, preprocessing_data, features)
\end{minted}
The Seer training script outputs the models as C++ headers which take as input the set of input features and outputs a classification. The classifier selection model takes as input the set of trivially known features and gives as output a classification between the models which are a member of its selectable classes. The individual kernel classification models take as input a feature vector which is determined by the size of the input features given at training time and give, as output, an index which maps into an output vector determining the output kernel.
\section{Case Study}
\label{sec:case_study}

We evaluated our framework, Seer, using Sparse-Matrix Vector Multiplication. SpMV is a fundamental sparse linear algebra routine and is representative of the dataflow that is common in graph analytics and other irregular workloads. SpMV can be performed using various sparse formats~\cite{Filippone:2017:SMM} and load balancing strategies~\cite{Osama:2023:APM,Merrill:2016:MPS,Dalton:2015:OSM}. These load balancing strategies and sparse formats are often combined into various different kernels which are amalgamations of both a compressed format and applied load balancing strategy. Table~\ref{tab:spmv_kernel_explanation} shows all of the different variants which we have considered for this work. For this section, we have chosen to compare against an Oracle predictor, which is the perfect case where the best kernel is selected after all kernels are run so the perfect kernel is always selected. We believe this serves as the best comparison for our work as it evaluates the predictor against its maximum possible capacity and shows the error in absolute terms versus the best possible performance it could achieve.

\begin{figure*}
    \begin{subfigure}[t]{0.48\textwidth}
        \begin{tikzpicture}[      
            every node/.style={anchor=south west,inner sep=0pt},
            x=1mm, y=1mm,
          ]   
         \node (fig1) at (0,0)
           {\includegraphics[width=\linewidth]{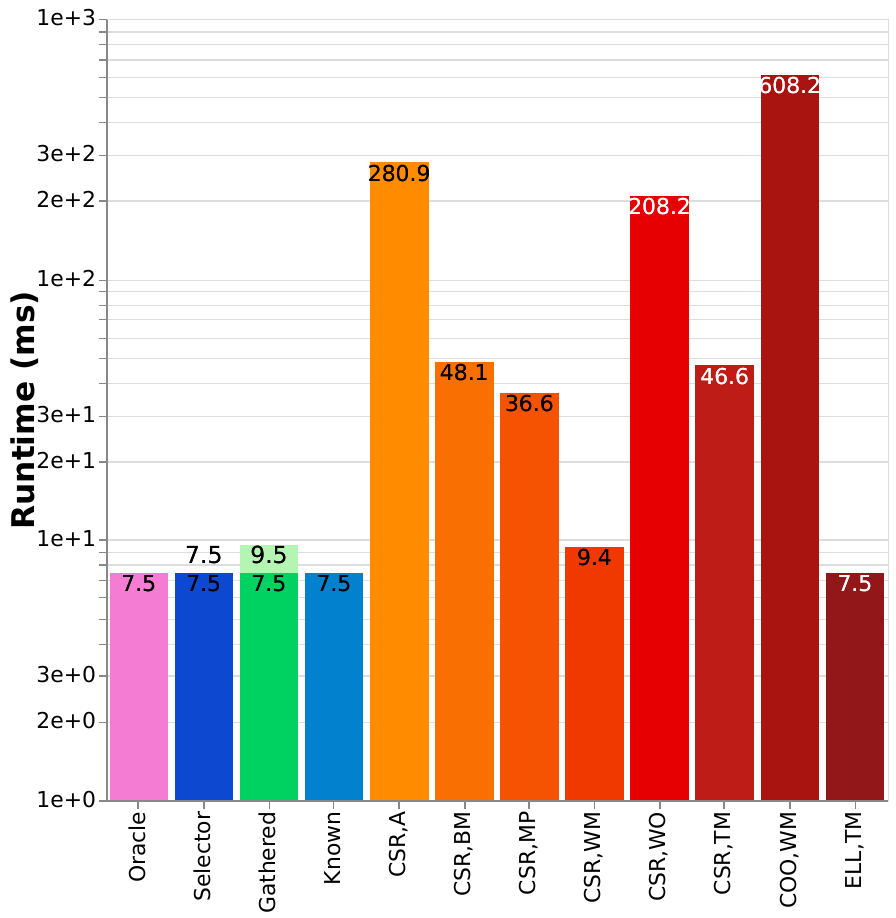}};
         \node (fig2) at (11,66.5)
           {\includegraphics[scale=0.16]{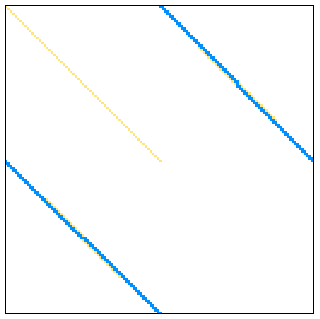}};  
        \end{tikzpicture}
        \caption{\textbf{nlpkkt200}, Classifier selection Known Predictor. \label{fig:nlpkkt200}}
    \end{subfigure}
    \hfill
    \begin{subfigure}[t]{0.48\textwidth}
        \begin{tikzpicture}[      
            every node/.style={anchor=south west,inner sep=0pt},
            x=1mm, y=1mm,
          ]   
         \node (fig1) at (0,0)
           {\includegraphics[width=\linewidth]{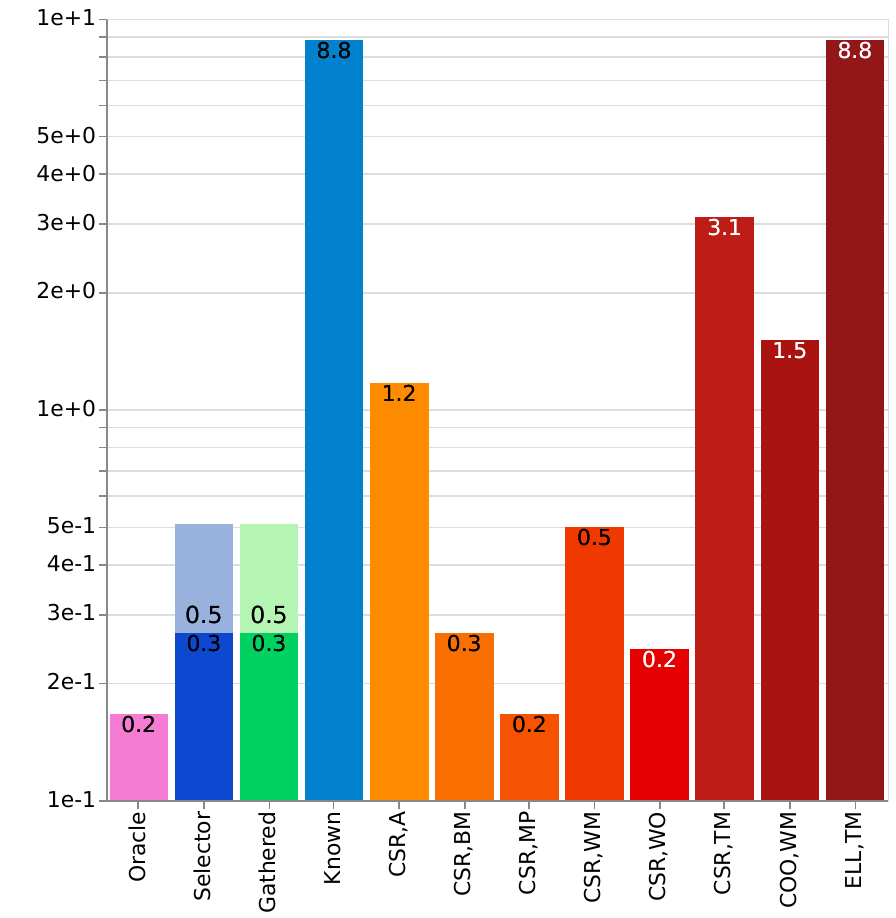}};
         \node (fig2) at (11,68.5)
           {\includegraphics[scale=0.10]{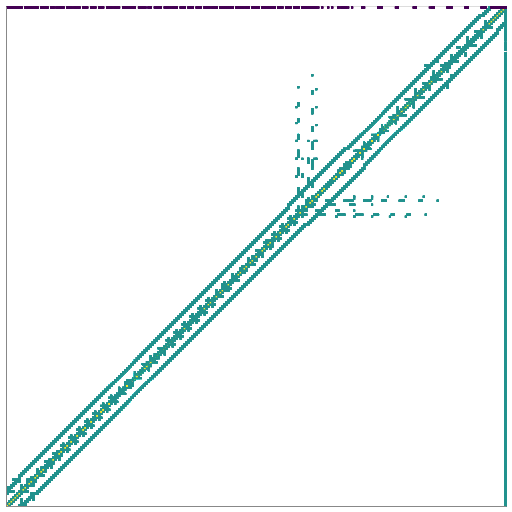}};  
        \end{tikzpicture}
        \caption{\textbf{matrix-new\_3}, Classifier selection prefers Gathered Predictor. \label{fig:matrixnew}}
    \end{subfigure}
    \vfill
    \begin{subfigure}[t]{0.48\textwidth}
        \begin{tikzpicture}[      
            every node/.style={anchor=south west,inner sep=0pt},
            x=1mm, y=1mm,
          ]   
         \node (fig1) at (0,0)
           {\includegraphics[width=\linewidth]{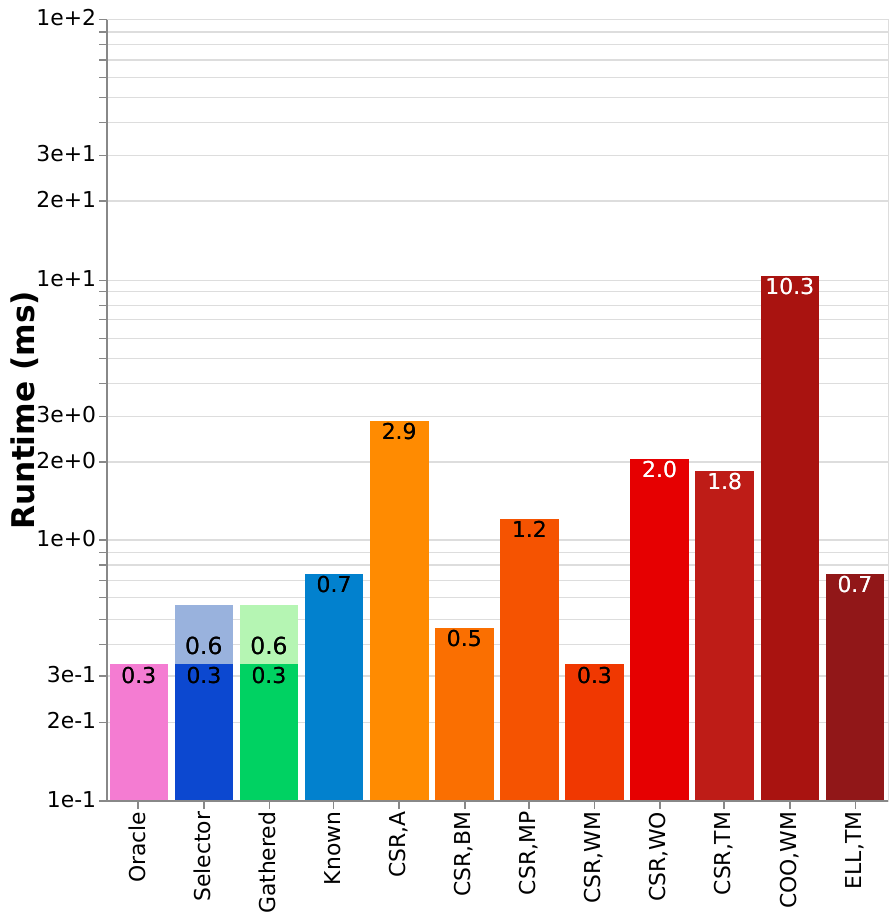}};
         \node (fig2) at (11,67)
           {\includegraphics[scale=0.11]{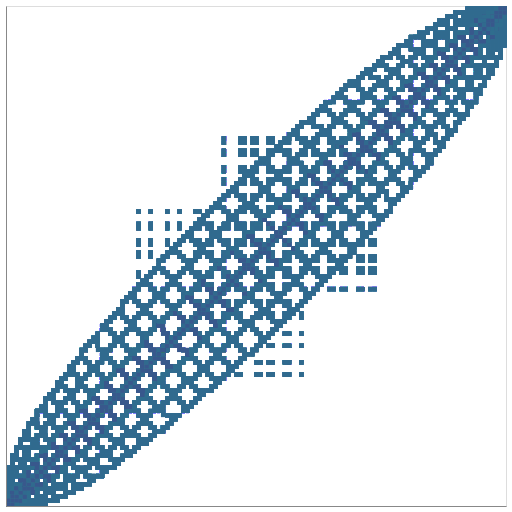}};  
        \end{tikzpicture}
        \caption{\textbf{Ga41As41H72}, Classifier selection prefers Gathered predictor. \label{fig:Ga41As41H72}}
    \end{subfigure}
    \hfill
    \begin{subfigure}[t]{0.48\textwidth}
        \includegraphics[width=\linewidth]{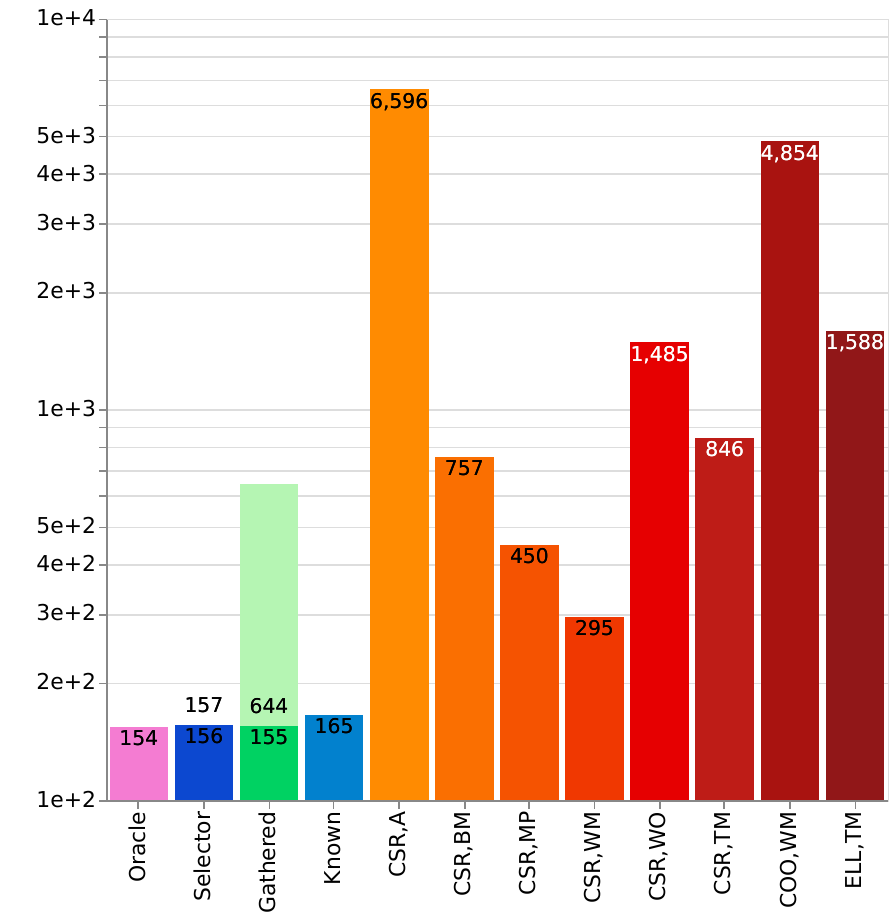}
        \caption{Aggregate SPMV single-iteration runtime across the dataset. \label{fig:aggregate}}
    \end{subfigure}

    \caption{Single iteration runtime for the Oracle versus the \textbf{Classifier Selection}, \textbf{Gathered},\ and \textbf{Known} Data predictors using sparse-matrices from SuiteSparse. Refer to Table~\ref{tab:spmv_kernel_explanation} for description of the label used to denote format and load-balancing schedule of the kernels. Lighter-stacked bars show overhead for a given approach. A visualization of the matrix is shown in the top left.
        Our classifier selection predictor achieves competitive performance against the speed-of-light Oracle kernel selection.
        \label{fig:single_iteration_plots}}
\end{figure*}


\begin{table*}
    \small
    \centering
    \caption{Various load balancing schedules and the sparse matrix formats used for the SpMV case study~\cite{Busato:2017:APP}.
        \label{tab:spmv_kernel_explanation}}
    \begin{tabularx}{\linewidth}{nbk}
        \toprule
        Load-Balancing Alg.                                                             & Description                                                                               & Format   \\
        \midrule
        Adaptive-CSR (A) and \textbf{rocSPARSE}~\cite{Daga:2015:SAS,AMD:2018:rocSPARSE} & Bins rows based on their row-sizes into small, medium, large, and processes them together. & CSR      \\
        Thread Mapped (TM)~\cite{Bell:2008:ESM}                                         & Maps one row or fixed width of the sparse matrix to a single thread.                      & CSR, ELL \\
        Warp Mapped (WM)~\cite{Bell:2008:ESM,Merrill:2012:SGG,Davidson:2014:WPG}        & Maps one or fixed number of rows of the sparse matrix to a wavefront.                     & CSR, COO \\
        Block Mapped (BM)~\cite{Brahmakshatriya:2021:CGA,Osama:2023:APM}
                                                                                        & Maps one row of the sparse matrix to a thread group.                                      & CSR      \\
        Work-Oriented (WO/MP)~\cite{Merrill:2016:MPS}                                   & Divides the total "work" (nonzeros + number of outputs) to each thread.                   & CSR      \\
        \bottomrule
    \end{tabularx}%

\end{table*}

For the model predicting based on trivially known features, we use metrics which accompany the input dataset, available at runtime, such as number of rows, columns, and nonzeros within a sparse-matrix. For the model predicting on dynamically computed features, it is important to select statistics which represent the way in which parallelism is being delineated to the GPU. Because our kernels generally utilize an inner-product formulation for SpMV, we implemented parallel statistic collection kernels and selected statistics which are representative of the row-order density characteristics of the plot in the form of max row density, minimum row density, variance of row density, and mean row density.

For some load balancing strategies it is necessary to first compute some preprocessing step. Often, this preprocessing step occurs only once over the span of iterations. In some libraries, this preprocessing cost is considered to be amortized in most cases. However, there is typically a crossover point where the cost of preprocessing gets amortized. One such example of this is the Adaptive-CSR implementation, in which before starting the problem, the rows within the matrix must be binned sequentially in several categories based on the size of the rows; these like-sized rows are then processed together in parallel on the GPU~\cite{Daga:2015:SAS}. The sequential binning preprocessing step has a large cost in the initial run, but gets amortized with a significant decrease in per-iteration runtime on large sparse matrices.

\subsection{Feature Collection \label{sec:metadata}}

One important distinction is between row-order statistics and row-order density statistics. The density characteristics are characteristics that are normalized to the overall size of a row. For example, given a row that is 20 elements long, five of which are nonzero, the density will be 0.25. This is distinctive of the number of nonzero in that row, which is five. This was selected as a way to normalize data across the entire dataset and to represent the feature as a metric of both problem size and row-size rather than one or the other.

\begin{figure}
    \includegraphics[width=\linewidth]{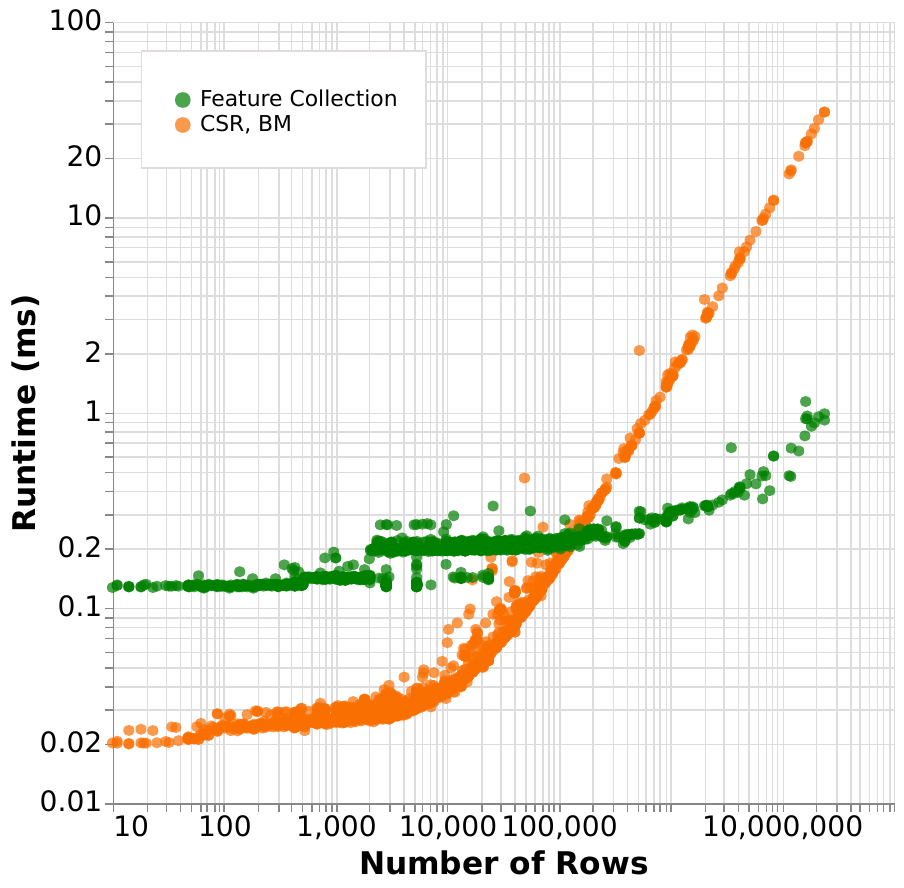}
    \caption{A plot showing the feature collection cost vs. the runtime of the CSR,BM Kernel for sparse-matricies with less than 10 Million Rows.\label{fig:metadata_collection_vs_rows}}
\end{figure}

To dynamically compute features, we have written parallel GPU kernels to loop over the offsets of a CSR representation of the sparse-matrix. Because we are parallelizing across the offsets, and the size of the offsets grows in scale with the number of rows, we can see a trend in which the cost of feature collection grows with the number of rows, as shown in Fig.~\ref{fig:metadata_collection_vs_rows}. On the left side of Fig.~\ref{fig:metadata_collection_vs_rows}, we see that the cost of feature collection is significant compared to the kernel runtime. However, after the 100,000 rows crossover point, the runtime increases with a different slope than the cost of feature collection. This shows effectively the ability of the feature collection to be worthwhile. However it is a problem dependent decision process.  The dynamically computed features collected for training are output to a file which is merged with the load balancing kernel performance characteristics before being ingested by the training process.

Given the row-order statistics, a domain expert may extract some information about the graph. We conjecture that the model will also be able to determine these emergent properties of compatibility between a dataset and load balancing strategy. For example, given the largest row in a dataset, someone with domain expertise can infer that this will be the "slowest thread" given a load balancing schedule which maps a row to each thread. The variance of the row density will be representative at a high level of the distribution of the data across rows, which will provide information on how the load imbalance may be structured across the matrix and may be combined with other statistics to determine an emergent structure of the data, without granular examination of the data.

Ideally, each feature will be representative of a characteristic that intuitively maps to the delination of parallelism to the device. By considering the selection of the dynamically computed features in this light, it provides a more clear framework of how the irregular parallelism may be mapped onto the regular lanes of the device. We propose that gathering these statistics in an intentional way for a workload allows the irregular parallelism to be expressed in a minimal format. It is immensely important to consider the cost of gathering a statistic and how the chosen statistic is representative of the sparsity's structure to provide the model with the ideal characteristics for selection of load balancing kernel.

\begin{table}
    \centering
    \caption{Shows the Kendall Correlation coefficient between load balancing strategies and the features dynamically computed from the data. Each load balancing kernel's performance has a different monotonic correlation with different matrix characteristics. Intuitively, these are relationships that are leveraged by a predictor. This method shows the importance of individual characteristics in the decision making process. \label{tab:correlation}}

    \begin{tabularx}{\columnwidth}{ccccccc}
        \toprule
        Load-Balancing Alg. & rows & nnz  & Most & Least & Avg  & Var  \\
        \midrule
        Adaptive-CSR (A)    & 0.88 & 0.67 & 0.40 & 0.77  & 0.54 & 0.50 \\
        CSR,BM              & 0.86 & 0.67 & 0.41 & 0.76  & 0.54 & 0.50 \\
        CSR,MP              & 0.69 & 0.63 & 0.30 & 0.60  & 0.42 & 0.39 \\
        CSR,WM              & 0.40 & 0.58 & 0.02 & 0.35  & 0.09 & 0.08 \\
        CSR,WO              & 0.66 & 0.80 & 0.29 & 0.61  & 0.32 & 0.31 \\
        CSR,TM              & 0.41 & 0.53 & 0.01 & 0.41  & 0.15 & 0.07 \\
        COO,WM              & 0.65 & 0.86 & 0.30 & 0.56  & 0.28 & 0.31 \\
        ELL,TM              & 0.24 & 0.43 & 0.13 & 0.22  & 0.03 & 0.07 \\
        \bottomrule
    \end{tabularx}%
\end{table}

The Kendall correlation coefficient is a method for determining if there is a montonic relationship between sets. Given the statistics mentioned above, we calculated the Kendall correlation coefficient of the runtime of individual kernels with the features provided to the various predictors and have included it in Table~\ref{tab:correlation}. For the Kendall correlation coefficient between the features and runtime, a larger value represents a strong relationship between the runtime and individual feature characteristics. Larger values for correlation imply that the values move tightly together, when one goes up or down the other often does too. For example, we see that CSR,WO is most correlated with the number of nonzero values, this is because CSR,WO takes a work oriented approach which extracts parallelism by splitting the work rather than the non zeroes. For the row parallelism based kernels we tend to see a higher value for the number of rows, which suggests that the runtime has a montonic relationship with the number of rows. This makes sense, as with more rows in a row-mapped load balancing strategy, there is more parallelism available, so we would expect a somewhat montonically decreasing relationship. However, due to load imbalance resulting from imbalance in the amount of work, the rows and runtime are not completely correlated. This nuance suggests that there is some relationship between the features and the underlying delineation of parallelism in the load balancing strategies which our predictors are able to exploit.

\subsection{Methodology}
For evaluation, we collected runtime metrics of various kernels described in Table~\ref{tab:spmv_kernel_explanation} on the entirety of SuiteSparse Matrix Collection dataset~\cite{Davis:2011:TUO} using AMD Instinct\textsuperscript{TM} MI100 accelerator. For Adaptive CSR implementation, we used the AMD sparse-linear algebra library rocSPARSE (version 5.6.0). All other kernels were implemented using The AMD ROCm\textsuperscript{TM} open software platform (version 5.6.0).
When collecting runtimes of different kernels, we deploy best practices of using 10 warm-up iterations and the reported runtime result averaged across 10 runs. Within Seer, we also provide parallel, GPU-accelerated subroutines to gather the relevant dynamically collected features for the sparse matrices as described in Section~\ref{sec:metadata} to minimize the cost of feature collection.

\subsection{Model Training and Accuracy }
For training, we utilize the Decision Tree Classifier implementation included in the scikit machine learning library. For training and accuracy computation, we use an 80-20 train-test set. On the test set the known, gathered, and classifier selection predictors were able to achieve accuracies of 77\%, 83\%, and 95\%, respectively.

For runtime use of these models, it is important to note the difference between runtime prediction accuracy and prediction error. Namely, the accuracy is calculated based on the exact correct guess of the fastest kernel. A low accuracy number would not necessarily imply that it is significantly less efficient than a single kernel. The accuracy is a measure of the number of mispredictions of a classifier model. This differs from error in that the error of the model is based on the total amount of runtime which is lost over the unachievable ideal Oracle approach. For the aggregate numbers in Fig. \ref{fig:aggregate} it is notable that the known predictor can perform very well in terms of error. However, a look at the underlying number of mispredictions shows that it is not due to its high accuracy, but rather the error of the mispredictions it does make is relatively low. In this way, the results display that there is some leniency between right and wrong in terms of error for our approach. In Fig.~\ref{fig:Ga41As41H72} two kernels may perform closely, such as CSR,BM and CSR,WM, which may produce an incorrect inference in absolute terms but does not produce a significant difference in the absolute runtime of the kernel.

\begin{figure*}[t]
    \centering
    \begin{subfigure}[t]{0.31\textwidth}
        \includegraphics[width=\linewidth]{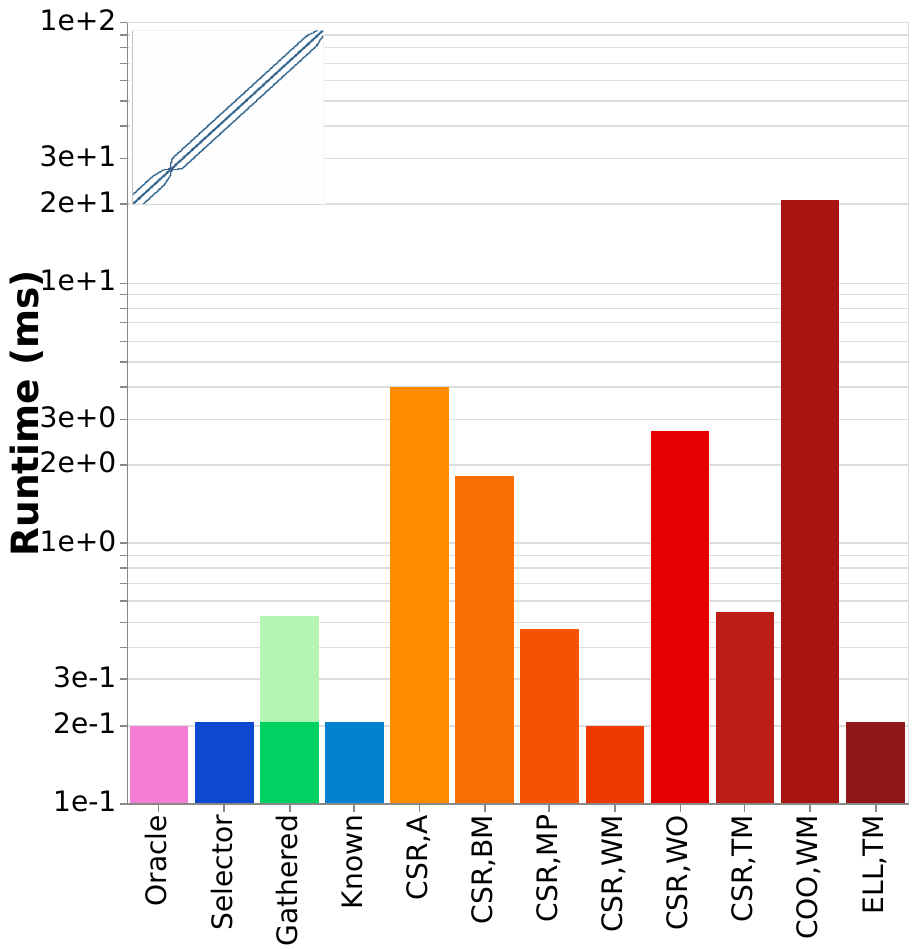}
        \caption{\textbf{CurlCurl\_3, 1 iteration}, Adaptive-CSR's preprocessing is not amortized and not selected. \label{fig:curlcurl_1iter}}
    \end{subfigure}
    \hfill
    \begin{subfigure}[t]{0.31\textwidth}
        \includegraphics[width=\linewidth]{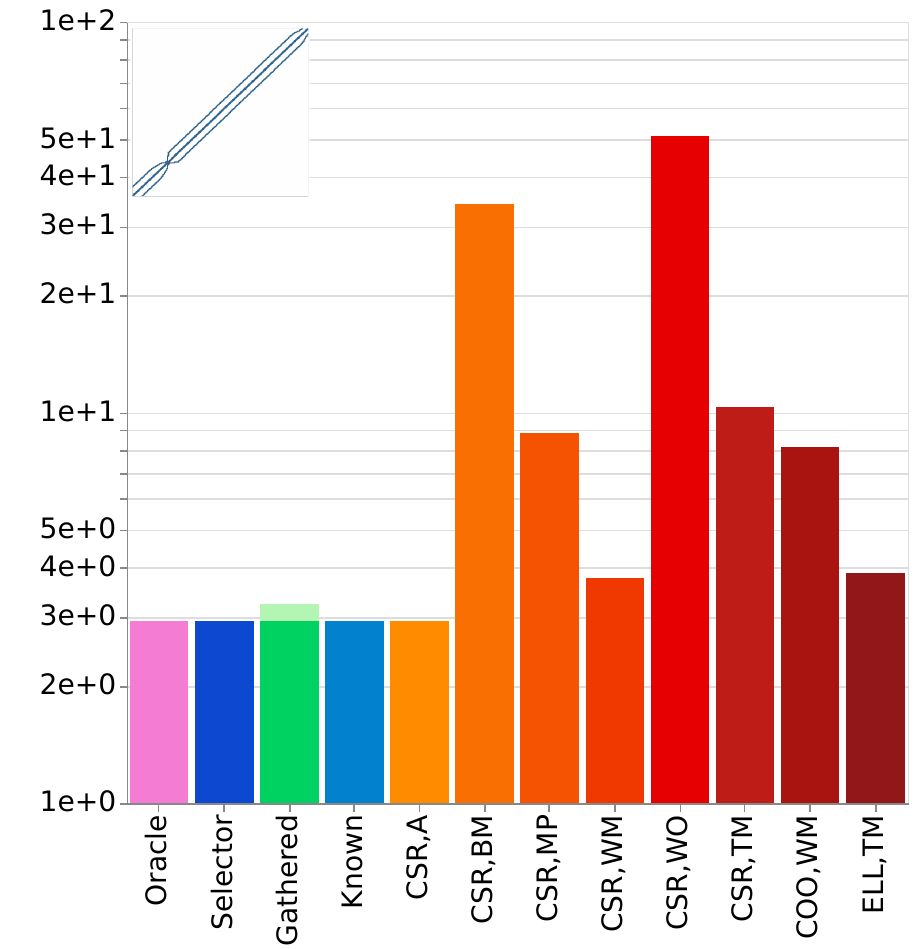}
        \caption{\textbf{CurlCurl\_3, 19 iterations}, Adaptive-CSR's preprocessing is amortized and selected. \label{fig:curlcurl_19iter}}
    \end{subfigure}
    \hfill
    \begin{subfigure}[t]{0.31\textwidth}
        \includegraphics[width=\linewidth]{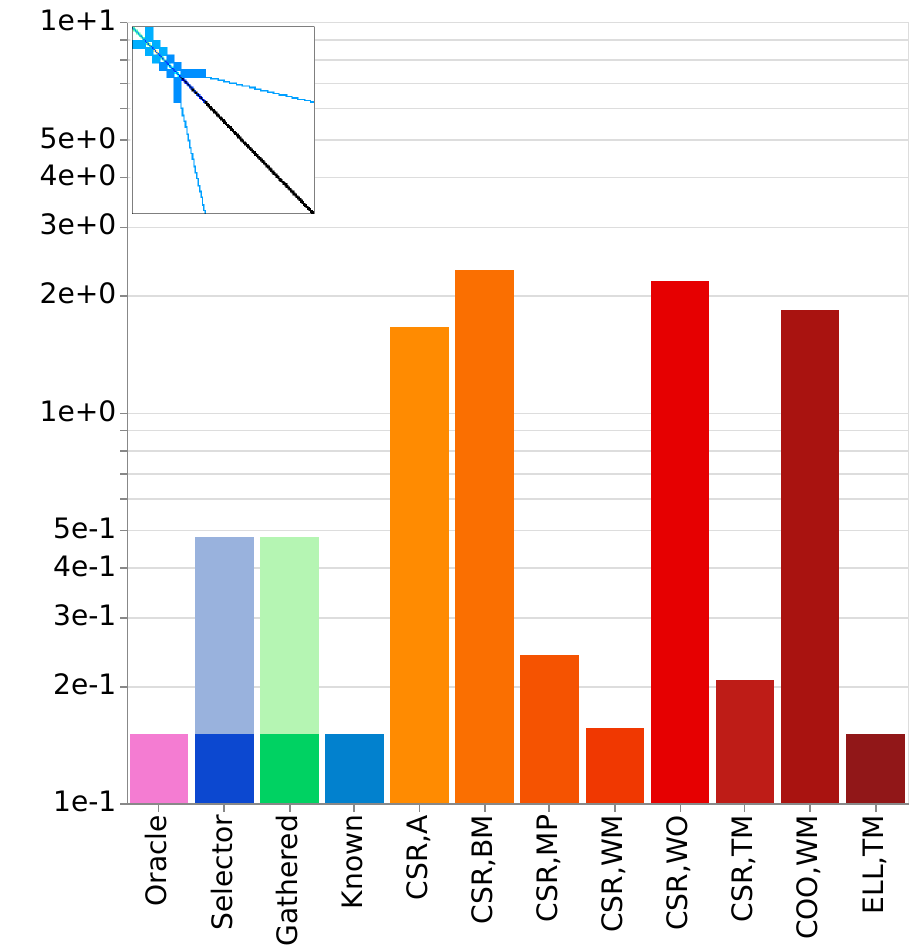}
        \caption{\textbf{G3\_Circuit, 1 iteration}, rocSPARSE preprocessing is not amortized and ELL,TM is selected. All predictors correctly predict.  \label{fig:g3_1iter}}
    \end{subfigure}
    \vfill
    \begin{subfigure}[t]{0.31\textwidth}
        \includegraphics[width=\linewidth]{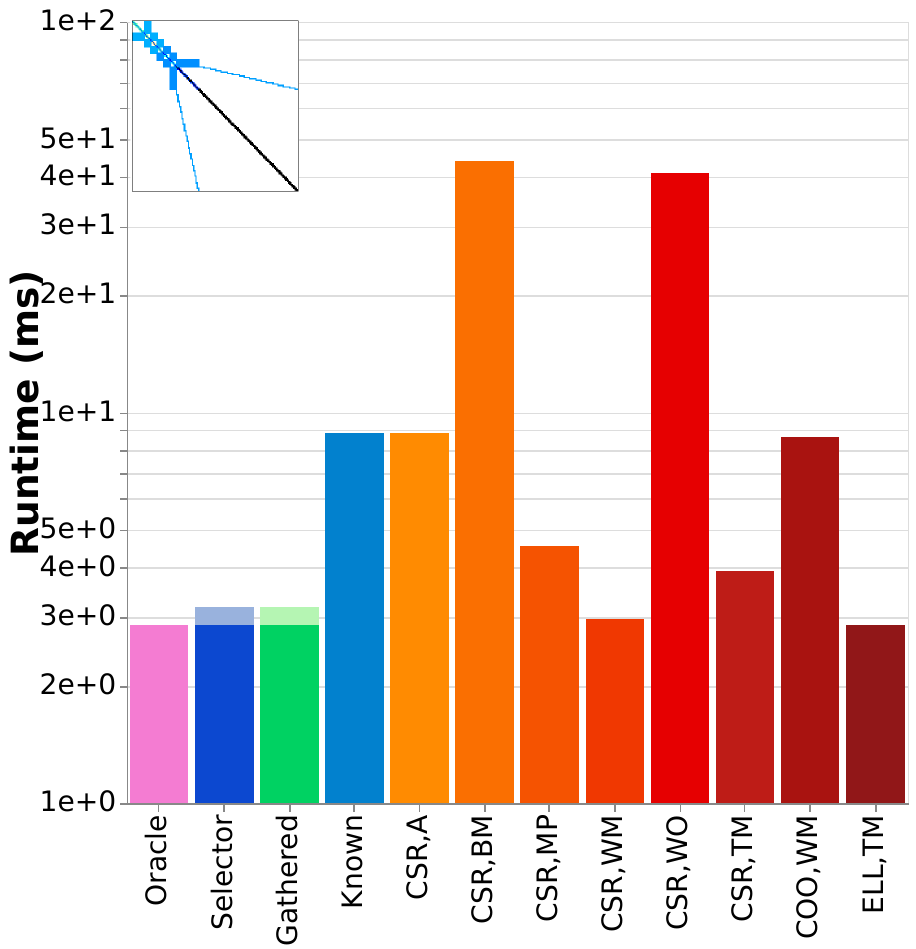}
        \caption{\textbf{G3\_Circuit, 19 iterations}, ELL,TM is still selected as rocSPARSE has not fully amortized the preprocessing cost, Known feature predictor is incorrect and predicts that the preprocessing will be amortized at this amount of computation.\label{fig:g3_19iter}}
    \end{subfigure}
    \hfill
    \begin{subfigure}[t]{0.31\textwidth}
        \includegraphics[width=\linewidth]{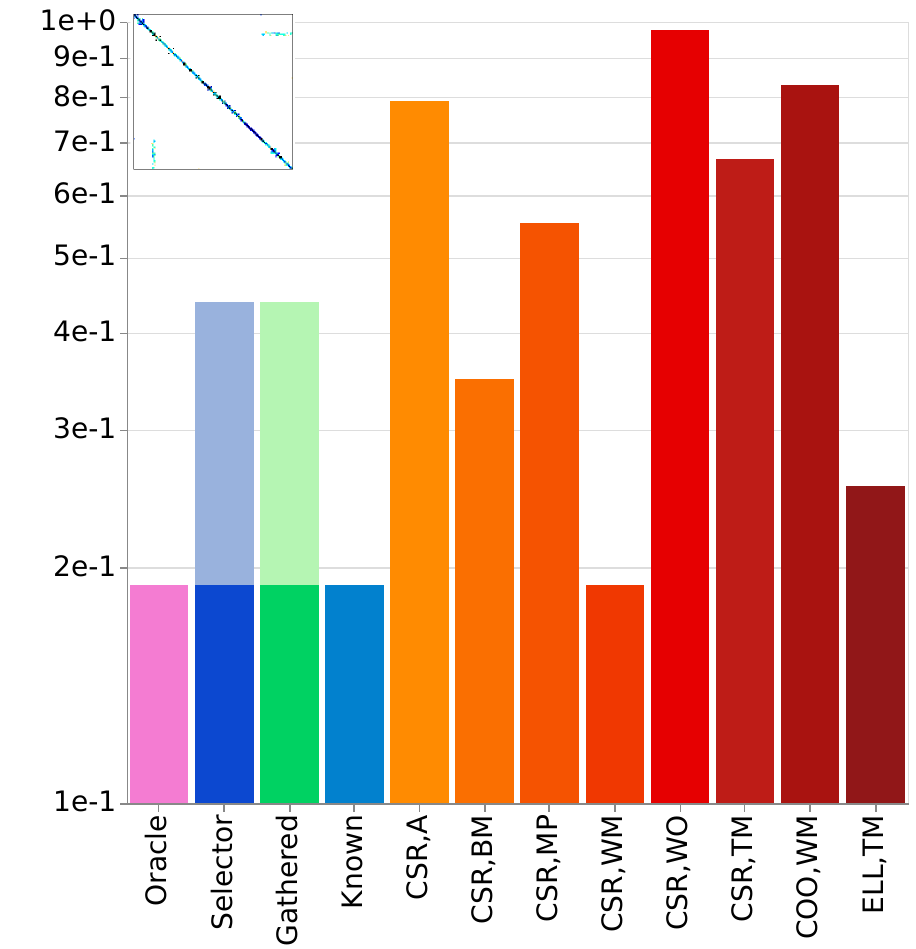}
        \caption{\textbf{PWTK, 1 iteration}, All predictors predict correctly. rocSPARSE Preprocessing is not amortized. \label{fig:pwtk_1iter}}
    \end{subfigure}
    \hfill
    \begin{subfigure}[t]{0.31\textwidth}
        \includegraphics[width=\linewidth]{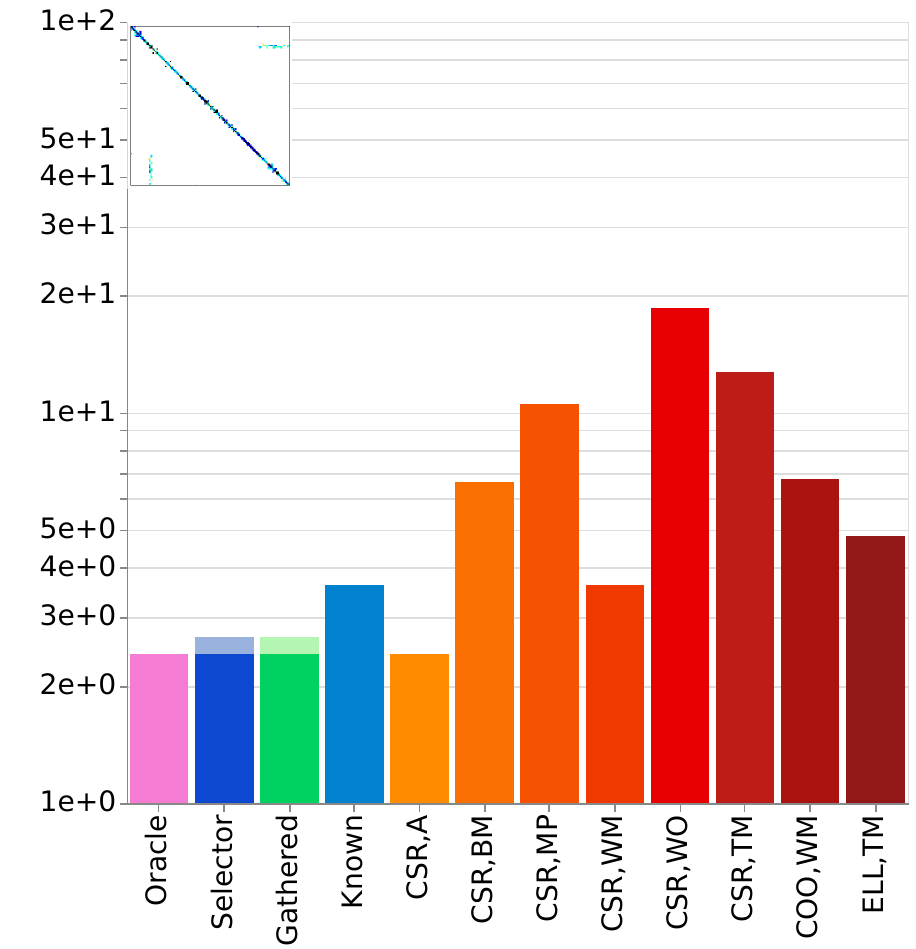}
        \caption{\textbf{PWTK, 19 iterations}, The gathered feature based predictor is able to predict amortization, whereas the known feature predictor is unable to correctly predict that rocSPARSE preprocessing will be amortized at this number of iterations.\label{fig:pwtk_19iter}}
    \end{subfigure}
    \caption{Multi-iteration analysis for model's ability to select amortized preprocessing kernels. \label{fig:multi_iteration}}
\end{figure*}

\subsection{Single Iteration Performance Evaluation}
For our case study, we examine both single and multiple-iteration SpMV kernels to accommodate for use cases such as iterative solvers, where SpMV is a core-routine. Single iteration SpMV is useful as it displays how the cost of feature collection may overcome the rest of the execution time. For multiple-iteration use cases, it is important for the model to be able to predict if the preprocessing of an individual kernel will be amortized. Fig.~\ref{fig:single_iteration_plots} shows both the aggregate runtime and some particularly interesting selected plots. In these plots we compare against the \emph{Oracle}, which refers to an unachievable ideal kernel selector by exhaustively selecting the fastest kernel for each problem. We also note that due to the use of a lightweight decision tree, the cost of inference is negligible but accounted for in our predictor.

In the results in Fig.~\ref{fig:nlpkkt200},~\ref{fig:Ga41As41H72}, and~\ref{fig:aggregate}, the feature collection cost significantly increases the overall cost of the gathered feature predictor. However, the gathered feature predictor is able to predict more accurately as seen in Fig.~\ref{fig:matrixnew} and~\ref{fig:Ga41As41H72}. This is an important nuance, as many works\cite{Muralidharan:NITRO}\cite{Yesil:WISE} which propose similar ideas do not take into account this feature collection cost, which can significantly increase the runtime. Instead, we see the classifier selection model electing to go for the known data model in most cases where its probability of success is high and it will be reasonably accurate. The classifier selection chooses to go for the gathered feature model, however, when the cost of a mispredict may be high. This results in the classifier selection model being able to avoid feature collection in most instances, but still being close in performance overall to the unachievable (at runtime) Oracle.

Fig.~\ref{fig:nlpkkt200} shows an example in which the feature collection cost increases the gathered predictor's runtime to be larger than any other predictor. The classifier selection predictor chose to go with the Known Data Predictor, which resulted in a lower runtime than if it went with the gathered data predictor. This is important to show as other works do not display the cost of feature collection. Thus this cost is invisibly inserted when the model is run.

Fig.~\ref{fig:Ga41As41H72} shows a case where collection cost was worthwhile even for a single iteration. In this case, the known data predictor lacked some amount of information which it needed to make an accurate inference. Due to this, it was unable to select the best kernel. The gathered feature predictor was able to select the best kernel, but its runtime was not the same as the minimum kernel overall due to the cost of the feature collection.

As shown in Fig.~\ref{fig:aggregate}, the classifier selection model is able to perform with 2$\times$ the performance of any individual kernel and 6.5$\times$ geomean speedup over all individual kernels.  Overall, the single iteration numbers address the feature collection cost while being able to achieve performance which is close to the impractical \emph{Oracle} predictor.

\subsection{Multiple Iteration Performance Evaluation}

Additionally, a predictor was trained on data which had various numbers of iterations. In this study, the predictor was often able to correctly predict the iterations threshold at which the cost of kernels with large preprocessing costs would be amortized. This ability is important, as this is another common characteristic that is available to all predictors at runtime. In addition, for work involving multiple iterations it is also more realistic to use the gathered data predictor, as the feature collection cost may also be amortized over the span of many iterations. These effects are shown in Fig.~\ref{fig:multi_iteration}. In Fig.~\ref{fig:curlcurl_1iter} we see CSR, Work Mapped being selected which does not include a preprocessing step. However, in Fig.~\ref{fig:curlcurl_19iter} Adaptive-CSR is selected which includes a preprocessing step which is amortized over the larger number of iterations. It was selected to display 19-iterations as this was the crossover point for some but not all of the graphs, which led to it being more difficult for the predictors to correctly predict if amortization of preprocessing costs would occur.


Fig.~\ref{fig:g3_1iter} and~\ref{fig:g3_19iter} show an example with multiple iterations, but the preprocessing cost of Adaptive-CSR is not amortized. In Fig.~\ref{fig:g3_1iter}, a single iteration of the G3\_Circuit sparse matrix is shown. In this example, the ELL,TM kernel is the fastest and all of the predictors are able to correctly predict that this is the fastest kernel. In Fig.~\ref{fig:g3_19iter} a multi-iteration run of the same sparse matrix is shown where the Known feature predictor predicts that the preprocessing cost of Adaptive-CSR will be amortized, but is incorrect due to its lack of information. This example shows that gathering dynamically computed features may be more worthwhile as it is another factor which may be amortized and will also allow the Gathered feature predictor to more accurately predict not only single iteration runs but also multi-iteration runs.

Fig.~\ref{fig:pwtk_19iter} shows an example in which initially the single iteration is correctly selected by all predictors, but the amortization of Adaptive-CSR's preprocessing step is not correctly predicted by the known feature predictor at 19 iterations.

Overall, the multiple iteration plots show the ability of the classifier selection model to correctly predict the amortization of preprocessing steps in kernels which require them. Additionally, the classifier selection model is able to outperform the amortized preprocessing steps alone across the entire dataset. As the classifier selection model is able to determine if the preprocessing will not be amortized at a particular number of iterations, it avoids the preprocessing cost when it determines it is unlikely to cause a costly misprediction. These results demonstrate that the predictor can detect the nuance of individual sparse matrices in the SuiteSparse dataset and correctly predict which kernel will be the most performant for a particular sparse matrix.

\section{Conclusion}
In this work, we show that it is possible to use our abstraction to select the best kernel at runtime and achieve performance close to the expensive \emph{Oracle} selector (i.e., running all implementations to select the best one). Our framework and abstraction addresses a complex problem of mapping sparse-irregular problems to the respective highest-performing kernel implementations. We also present our work as a generalized abstraction extensible to other kernels and algorithms using a simple API.

In future works, we hope to investigate classifier selectors of more feature collection strategies, alternate case studies, and the correlation of different statistics to the best-performing kernels. We aim to use the \emph{Seer} to develop better understanding  of the tradeoffs between different feature collection strategies and apply our framework to study other techniques used within the irregular-problem space.

\flushend
\begin{flushright}
\bibliographystyle{IEEEtran} 
\bibliography{loops_predict}
\end{flushright}


\end{document}